\documentclass[aip,amsmath,amssymb,reprint]{revtex4-2}
\usepackage{graphicx}
\usepackage{dcolumn}
\usepackage{bm}
\usepackage[utf8]{inputenc}
\usepackage[english]{babel}
\usepackage[T1]{fontenc}
\usepackage{mathptmx}
\usepackage{hyperref}

\usepackage{letltxmacro}
\LetLtxMacro{\ORIGselectlanguage}{\selectlanguage}
\makeatletter
\DeclareRobustCommand{\selectlanguage}[1]{%
	\@ifundefined{alias@\string#1}
	{\ORIGselectlanguage{#1}}
	{\begingroup\edef\x{\endgroup
			\noexpand\ORIGselectlanguage{\@nameuse{alias@#1}}}\x}%
}
\newcommand{\definelanguagealias}[2]{%
	\@namedef{alias@#1}{#2}%
}
\makeatother
\definelanguagealias{en}{english}
\definelanguagealias{EN}{english}

\draft 

\begin{document}
\title{Ultra-low-vibration closed-cycle cryogenic surface-electrode ion trap apparatus}

\author{T. Dubielzig}
	\email[]{dubielzig@iqo.uni-hannover.de}
	\affiliation{Institut f\"ur Quantenoptik, Leibniz Universit\"at Hannover, Welfengarten 1, 30167 Hannover, Germany}
\author{S. Halama}
	\affiliation{Institut f\"ur Quantenoptik, Leibniz Universit\"at Hannover, Welfengarten 1, 30167 Hannover, Germany}
\author{H. Hahn}	
	\affiliation{Physikalisch-Technische Bundesanstalt, Bundesallee 100, 38116 Braunschweig, Germany}
	\affiliation{Institut f\"ur Quantenoptik, Leibniz Universit\"at Hannover, Welfengarten 1, 30167 Hannover, Germany}		
\author{G. Zarantonello}
	\affiliation{Institut f\"ur Quantenoptik, Leibniz Universit\"at Hannover, Welfengarten 1, 30167 Hannover, Germany}		
	\affiliation{Physikalisch-Technische Bundesanstalt, Bundesallee 100, 38116 Braunschweig, Germany}
\author{M. Niemann}
	\affiliation{Institut f\"ur Quantenoptik, Leibniz Universit\"at Hannover, Welfengarten 1, 30167 Hannover, Germany}
\author{A. Bautista-Salvador}
	\affiliation{Physikalisch-Technische Bundesanstalt, Bundesallee 100, 38116 Braunschweig, Germany}
	\affiliation{Institut f\"ur Quantenoptik, Leibniz Universit\"at Hannover, Welfengarten 1, 30167 Hannover, Germany}
\author{C. Ospelkaus}
	\affiliation{Institut f\"ur Quantenoptik, Leibniz Universit\"at Hannover, Welfengarten 1, 30167 Hannover, Germany}
	\affiliation{Physikalisch-Technische Bundesanstalt, Bundesallee 100, 38116 Braunschweig, Germany}

\date{\today}

\begin{abstract}
We describe the design, commissioning and operation of an ultra-low-vibration closed-cycle cryogenic ion trap apparatus. One hundred lines for low-frequency signals and eight microwave / radio frequency coaxial feed lines offer the possibility of implementing a small-scale ion-trap quantum processor or simulator. With all supply cables attached, more than 1.3\,W of cooling power at 5\,K is still available for absorbing energy from electrical pulses introduced to control ions. The trap itself is isolated from vibrations induced by the cold head using a helium exchange gas interface. The performance of the vibration isolation system has been characterized using a Michelson interferometer, finding residual vibration amplitudes on the order of 10\,nm rms. Trapping of $^9$Be$^+$ ions has been demonstrated using a combination of laser ablation and photoionization. 
\end{abstract}


\maketitle

\section{Introduction}
Trapped atomic ions are a well controlled quantum system with applications in scalable quantum logic~\cite{monz_realization_2016,debnath_demonstration_2016,wineland_nobel_2013}, frequency metrology~\cite{ludlow_optical_2015}, quantum simulation~\cite{schaetz_focus_2013} and other areas. While ground state cooling of the quantized motional degrees of freedom of trapped ions~\cite{diedrich_laser_1989,monroe_resolved-sideband_1995} has been a key enabling step for applications, the traps themselves can be kept at room temperature. Nevertheless, cryogenic cooling of the ion trap itself to below 30\,K can represent an important advantage~\cite{chiaverini_insensitivity_2014}. In a cryogenic vacuum apparatus, the achievable pressures can be much lower than for room-temperature ultra-high vacuum systems, enabling long lifetimes of the trapped particles. This can be of particular importance in the case of small-scale micro-fabricated and surface-electrode traps~\cite{chiaverini_surface-electrode_2005, seidelin_microfabricated_2006} used in the context of quantum computing and simulations, as they may exhibit relatively shallow potential well depths, making the ion(s) susceptible to loss from collisions with background gas atoms. Even before ion loss occurs, collisions could affect the motional state of the ion(s), which is detrimental in many applications. In frequency metrology, an important systematic effect, namely the black-body radiation shift of atomic clock transitions due to the finite temperature of the ion's thermal environment, can be strongly suppressed by keeping the ion in a cryogenically cooled environment~\cite{poitzsch_cryogenic_1996}. Finally, electric field noise from the trap electrodes may excite the ions' motion in an uncontrolled way. This effect, known as `anomalous motional heating'~\cite{turchette_heating_2000}, can be an important systematic effect in ion clocks through uncontrolled Doppler shifts~\cite{ludlow_optical_2015}, and may affect the fidelity of multi-qubit quantum logic gates~\cite{brownnutt_ion-trap_2015}. This effect appears to be thermally activated and can thus be strongly suppressed by cooling the trap electrodes to cryogenic temperatures~\cite{deslauriers_scaling_2006,labaziewicz_suppression_2008,chiaverini_insensitivity_2014}. 

A potential concern for cryogenic operation is the added experimental overhead. Compared to other quantum systems, however, which may be impossible to operate without deep cryogenic cooling to $\approx\mathrm{mK}$ temperatures, ions already achieve the above advantages with moderate cooling to below 10\,K. Another concern may be the consumption of liquid helium, in particular since many ion-trap applications require optical access for laser beams, which increases the heat load on the system and the liquid helium consumption. An obvious solution, the use of a closed-cycle refrigeration system, typically introduces issues due to mechanical vibrations from the cryo cooler itself, which can affect the performance of the system, in particular with respect to focused-laser-beam interrogation and manipulation of the ion(s). The limited cooling power requires a careful design of the vacuum system and cryogenic shields to minimize the effect of thermal radiation and of the electrical connection system because of the potential introduction of heat through these connections. 

Here we present a cryogenic ion trap apparatus for surface-electrode ion traps based on a closed-cycle refrigeration system with minimal mechanical vibrations. We describe the experiment concept and the surface-electrode trap. We discuss the electrical connection system, the vacuum system, vibration isolation system, laser sources and imaging system and finally show successful trapping of $^9$Be$^+$ ions.

\section{Concept}

\subsection{Design goals}

\begin{figure}
		\centering
		\includegraphics{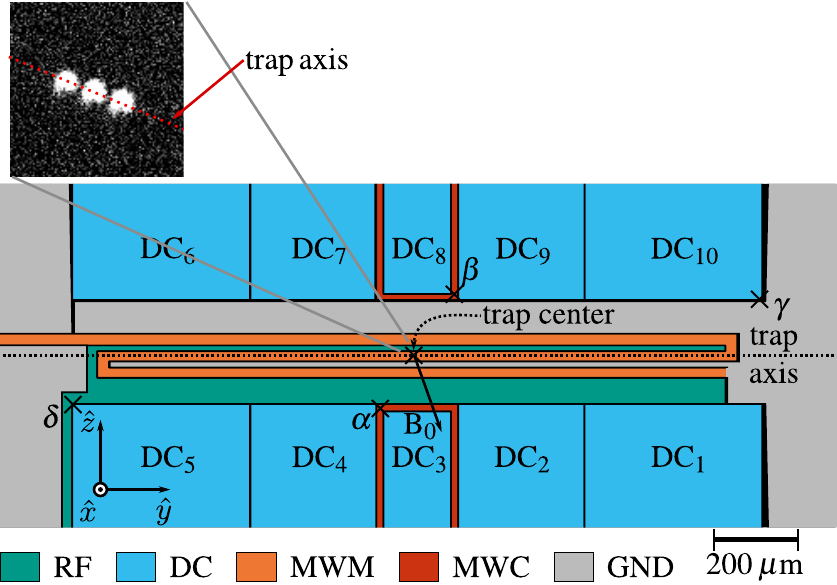}
		\caption{Trap electrodes around the trap center (see main text). In the top left, a picture of the resonance fluorescence of three trapped $^9$Be$^+$ ions is shown. Points $\alpha$, $\beta$, $\gamma$ and $\delta$ are used for laser beam alignment (see section~\ref{sec:trap_loading}). B$_0$ is the direction of the quantization axis, provided by a set of magnetic-field coils (see section \ref{sec:trap_loading}).}
		\label{fig:trap_electrodes}
\end{figure}

The application scenario of the setup discussed here is for scalable quantum computing and simulations with trapped ions using microwave-frequency magnetic field gradients for the underlying multi-qubit interactions. The microwave fields at the frequency of a field-independent hyperfine qubit in $^{9}\textnormal{Be}^{+}$ are generated by currents in trap-embedded conductors~\cite{ospelkaus_trapped-ion_2008} and directly address that transition and the associated motional sidebands. Design \cite{carsjens_surface-electrode_2014}, characterization \cite{wahnschaffe_single-ion_2017}, two-qubit gates~\cite{hahn_integrated_2019} and high-fidelity operation~\cite{zarantonello_robust_2019} have been shown for a room temperature setup. Present limits to gate fidelities in that setup are related to experimental complications from finite lifetime and (re-)loading effects among others. Anomalous motional heating may eventually limit achievable gate fidelities, making it desirable to operate the setup at cryogenic temperatures. As far as trap and embedded microwave conductors design is concerned, cryogenic operation will change the conductivity, dielectric loss and permittivity of the involved materials; all of these effects have been taken into account in the trap design. Figure~\ref{fig:trap_electrodes} shows the trap geometry developed for initial trapping experiments, an adapted version of the geometry used in~\cite{hahn_integrated_2019,zarantonello_robust_2019}. Ions can be confined at a distance of $x=70\,\mu\mathrm{m}$ from the surface using an RF (radio frequency) potential applied to the conductor depicted in green and labeled `RF'; suitable potentials on electrodes $\mathrm{DC}_{1-10}$, which are blue and labeled `DC' (direct current) provide confinement along $\hat{y}$. The conductors shown in orange (labeled `MWM' -- `microwave meander') and red (labeled `MWC' -- `microwave carrier'), when operated at a currents near-resonant with an atomic transition, provide microwave fields suitable for multi- and single-qubit operations, respectively. The gray area that is labeled `GND' (ground) is the ground plane. Note that the s-shaped MWM electrode is grounded at its end.

\begin{figure}
	\centering
	\includegraphics{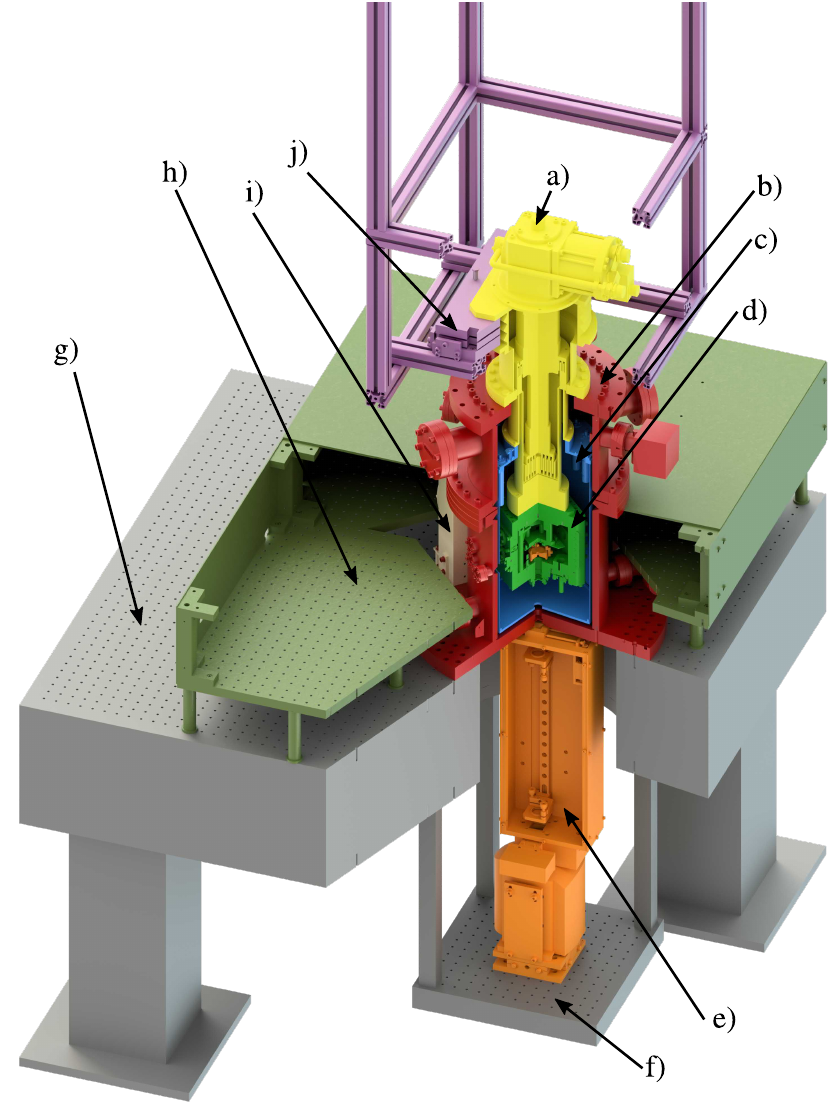}
	\caption{Overview of the functional components of the system. a) cold head with ultra-low-vibration interface, b) outer vacuum chamber, c) radiation shield, d) inner vacuum chamber, e) imaging system, f) lower breadboard, g) optical table, h) upper breadboard, i) magnetic field coils, j) cold head mount and translation stage.}
	\label{fig:whole_system_cut}
\end{figure}

The apparatus described here was designed to house and cool such a trap down to below 10\,K and provide XHV (extremely high vacuum) conditions. Laser access from two orthogonal directions was required as well as optical access for imaging of the trapped-ion resonance fluorescence. The design includes electrical connections for low-frequency and high-frequency currents and voltages. Furthermore, it is desirable to limit mechanical vibrations of the setup to allow for control with pairs of laser beams that need to exhibit interferometric stability at the position of the ions. If the trap is vibrating with respect to the table on which the laser optics are placed, the ions also exhibit that vibration and are moving with respect to the laser beams. For this reason, a vibration isolation is required. Such beams are used e.g. for ground state cooling of the ion(s). Figure~\ref{fig:whole_system_cut} shows a false color picture of the entire system which will be discussed in this paper.

\subsubsection{General cryostat design considerations}
The coldest part of a cryostat to which the experiment or sample is attached is called the cold finger. The design goal was to achieve between 3 and 10\,K at the cold finger, as further cooling typically only increases the complexity without significantly increasing the performance in the anticipated applications. An outer vacuum chamber eliminates convective and conductive heat loads, other than the conductive heat loads associated with the mechanical mounting of the cold finger, as well as cryo-condensation. A suitable isolation vacuum is typically below  $10^{-4}$\,mbar. A radiation shield between the outer vacuum chamber and the cold finger will shield the latter from room temperature black body radiation. A typical radiation shield temperature would be between 30\,K and 100\,K. To take full advantage of the XHV conditions provided by a cryogenic environment, a cold vacuum chamber containing the trap can be mounted to the cold finger. It has to be evacuated and sealed before cool-down and relies entirely on cryo pumping to maintain XHV conditions during operation of the system.

\subsubsection{Choosing the type of cooler}
\label{choosing_cryo}
In the required temperature range, one generally has the choice between a closed and an open system cryostat design. An example of an open system is a bath cryostat, with reservoirs of liquid nitrogen at 77\,K and liquid helium at 4.2\,K attached to the radiation shield and cold finger, respectively. Another open system is a flow cryostat in which the cooling liquids are continuously supplied from a reservoir. Both types of open systems boil off helium which can in part be recovered to lower overall costs. The constant need to refill a reservoir, the unpredictable costs due to changing helium prices and the lack of on-site available helium recovery systems led us to opt for a closed-cycle system. Prominent types of closed systems are Gifford-McMahon (GM) coolers and pulse tubes~\cite{ekin_experimental_2015}. 

Mechanical motion in both types of coolers can introduce significant amounts of vibrations. The vibration levels observed for GM coolers on the order of 100 $\mu$ms~\cite{ekin_experimental_2015} would already affect the stability of simple Doppler cooling and resonance fluorescence detection protocols, given typical laser beam waists of order $20\,\mu\mathrm{m}$. Pulse tube coolers can exhibit vibration levels up to two orders of magnitude lower than GM coolers~\cite{ekin_experimental_2015}. However, in many ion trap setups, phase-coherent pairs of laser beams are used to drive stimulated-Raman transitions for ground state cooling and quantum gates. Interferometric stability at the ion position would then be highly desirable. 
To give a specific example, references~\citenum{wang_vibration_2010} and \citenum{antohi_cryogenic_2009} demonstrate two-qubit laser-driven quantum gates at vibration levels of approximately $0.16 \lambda$, where $\lambda$ is the wavelength of the laser beam. In the case of $^9$Be$^+$ and for $\lambda = 313$\,nm, this would correspond to a vibration level on the order of 50\,nm.

For our application scenario, quantum gates are carried out using microwave fields that originate from trap-integrated conductors and are therefore not as sensitive to vibrations as laser-driven gates. Still, Raman beams are desirable for ground state cooling even in our scenario~\cite{hahn_integrated_2019,zarantonello_robust_2019}. Neither closed cycle system provides the desirable low vibration levels on their own. Different commercially available low-vibration interfaces for both types of coolers claim to achieve comparable performance levels that satisfy the requirements of interferometric stability. Arguments to consider in system design are the typically lower initial cost of GM coolers. A GM cooler, however, requires more frequent maintenance because of abrasion of moving cold-head parts, approximately every 10000\,h of operation vs. approximately every 20000\,h for a pulse tube cooler. The normal maintenance procedure of a GM cooler can be done on-site while a pulse tube cooler often requires shipping to the manufacturer. To reduce potential downtimes on the experiment because of cryo cooler maintenance, we opted for a GM cooler.

\section{Electrical connections}

\subsection{Electrical requirements}
Our application requires at least 2 high-frequency feed lines for microwave currents at about 1.08\,GHz with a few watts of peak power each, one for single-qubit operations and the other one for multi-qubit gates and interactions. To address different transitions in the hyperfine manifold, frequencies between 800 and 1500\,MHz are needed. We also require 1 line for an rf voltage at about 100\,MHz that connects to a step up resonator, which produces around 100\,V of amplitude applied to the RF electrode to ensure radial confinement of the ions. There are 10 control electrodes $\mathrm{DC}_{1-10}$ in this trap driven by low-frequency lines carrying voltages for axial confinement, micromotion compensation~\cite{berkeland_minimization_1998} and other radial and axial positioning of the ions. On these lines, mostly static voltages are applied, with transient currents occurring for ion transport due to the series RC character of the load. We use an in-vacuum cryogenic Schwarzschild-type objective for ion and trap imaging. For ease of alignment and focusing, 3D positioning of the objective is highly desirable. Each axis requires a piezo driven translation stage and each stage requires 2 current-carrying lines, which requires 6 low-resistance low-frequency lines in total. To be able to scale traps to a large number of ions (and electrodes) for applications in quantum computing and simulations, we opted for a design with 8 high-frequency and 100 low-frequency lines, 10 of which are low-resistance. Surface-electrode ion traps with small ion-to-electrode distances may operate with only a few volts on the low-frequency control electrodes, to be compared to hundreds of volts for macroscopic Paul traps. Quantum gates may require secular frequencies to be stable at the level of $10^{-6}$; spurious signal pickup from ground loops and other sources can therefore have a much more pronounced effect in these small traps. We have chosen to put the reference ground on the inner vacuum chamber and to have all externally attached signal sources, including high-frequency sources, referenced to that point and not grounded in any other way. 

\subsection{Electrical Design}

\subsubsection{Low-frequency lines}
Following the empirical Wiedemann-Franz law~\cite{franz_ueber_1853}, a low thermal conductance of cables (desirable to limit the static heat load introduced by an electrical connection) would typically go hand in hand with a high electrical resistance. The latter may be undesirable because of Joule heating if currents are to be run through the cable. For low-frequency signal lines attached to control electrodes, only transient currents will occur when voltages are changed to displace ions. Therefore, for 90 of the 100 low-frequency lines, we use varnished constantan wire with an outer diameter of 0.224\,mm (GVLZ087, GVL Cryoengineering). Compared to copper, constantan has a fairly high electrical resistivity but low thermal conductivity. Ten low-resistance low-frequency connections are realized using electrolytic copper wire of outer diameter 0.28\,mm (GVLZ175, GVL Cryoengineering). Without varnish, the diameters are 0.193\,mm and 0.25\,mm, respectively. The thermal conductivity integral is defined as the heat conduction $\dot{q}_{cond}$ through a solid of uniform cross section $A$ and length $l$ between two temperatures $T_1$ and $T_2$ with a temperature-dependent thermal conductivity $\lambda(T)$:
\begin{equation}
\dot{q}=\frac{A}{l} \int_{T_1}^{T_2} \lambda(T)
\label{thermal_conductivity_integral}
\end{equation}
Ekin~\cite{ekin_experimental_2015} provides these integrals between 4\,K and temperatures up to 300\,K for several materials. We used that data for constantan ($\lambda_{const.}(T)$) but chose the analytical thermal conductivity model provided by NIST Cryogenic Technology Resources, Index of Material Properties~\cite{noauthor_cryogenic_nodate} for copper, the main reason being that it provides data for a few more different copper purities. We chose RRR=100 (Residual Resistance Ratio) as a reference as this comes closest to the used wire's purity. According to the NIST data, one has:
\begin{equation}
\lambda_{RRR100}=10^{\frac{2.2154 -0.88068 \tilde{T}^{0.5} +0.29505 \tilde{T} -0.04831 \tilde{T}^{1.5} +0.003207 \tilde{T}^{2}}{1 -0.47461\tilde{T}^{0.5} +0.13871 \tilde{T} -0.04831 \tilde{T}^{1.5} +0.0012810 \tilde{T}^{2}}}\frac{\textnormal{W}}{\textnormal{m} \cdot \textnormal{K}},
\label{thermal_conductivity_copper1000}
\end{equation}
where $\tilde{T}$ is the temperature in kelvin. Assuming an ambient temperature of 296\,K, a radiation shield at 40\,K and an inner chamber at 4\,K, and also assuming perfect thermal anchoring of all wires at the respective temperature stages, with a wire length of 160\,mm between 296\,K and 40\,K as well as 260\,mm between 40\,K and 4\,K, we can calculate heat loads on both cold stages. One constantan wire produces a heat load of $\approx0.89$\,mW on the 40\,K stage and $\approx 0.03$\,mW on the 4\,K stage. One copper wire leads to a heat load of $\approx 49.1$\,mW on the 40\,K stage and $\approx 13$\,mW on the 4\,K stage. All 90 constantan wires combined lead to an insignificant heat load of $\approx 79.8$\,mW on the 40\,K stage and $\approx 3.1$\,mW on the 4\,K stage. All 10 copper wires contribute $\approx 491$\,mW to the 40\,K heat load and  $\approx 130$\,mW to the 4\,K heat load. The former can be easily absorbed by the more than 50\,W available cooling power at 40\,K but the latter consumes nearly 10\% of the available cooling power of about 1.5\,W at 4\,K.

We have applied the low-frequency lines in the following way: After cutting the wires to length, we removed the varnish by means of a motorized rotating blade wire stripper, with three quickly rotating blades with adjustable distance between them (ISOL-EX Typ 02, Fritz Diel GmbH \& Co. KG). We assembled a 25-pin D-sub UHV (ultra-high vacuum) compatible cable to provide two copper and 23 constantan wires. We used materials supplied by Allectra GmbH. One female PEEK (polyether ether ketone) connector (212-PINF-25-S) was mounted on each end. We folded the tip of each wire once by 180$^{\circ}$ before crimping it into a small crimp pin (212-PINM-25-S) because the wire gauge would be too small for the pin's inner diameter otherwise. The wires were then wound 10 times around and glued onto an 8\,mm diameter copper bobbin which was anchored to the 40\,K stage. We used a thermally conductive two-component epoxy with low outgassing properties: the epoxy resin LOCTITE$^{\textregistered}$ STYCAST 2850FT together with the catalyst LOCTITE$^{\textregistered}$ CAT 24LV. The 160\,mm long end was connected to a D-sub feedthrough on a DN63CF flange at the outer chamber and the 260\,mm long end was connected to a D-sub feedthrough welded into a custom flange at the inner chamber. The wiring between bobbin and inner chamber took place inside the radiation shield. A hole in the radiation shield, big enough for a 25 pin connector to pass through, was later mostly covered with a copper sheet, leaving a hole only large enough for the wires, so room temperature thermal radiation would not pass through.

\subsubsection{High-frequency microwave current lines}

For microwave frequencies between 800 and 1500\,MHz, coaxial cables are required to provide a well-defined impedance and shielding from noise sources. The same considerations as for the low-frequency lines apply here, except that the primary purpose of the microwave lines is to apply high currents inside the ion trap~\cite{ospelkaus_trapped-ion_2008}, which makes Joule heating a concern. In addition, for the frequency range considered, currents will not flow in the entire metallic conductor due to the skin effect, but only in a surface layer. The remaining metal does not contribute to electrical conduction, but introduces additional heat load through thermal conduction. As an acceptable compromise between expected Joule heating and low thermal conduction, we chose UT-141B-SS from Microstock. This is a semi-rigid coaxial cable with polytetrafluoroethylene (PTFE) as a dielectric between a BeCu center and a 304 stainless steel outer conductor. It has no isolation on the outside, which is beneficial for thermal anchoring. 

\begin{figure}
	\centering
	\includegraphics{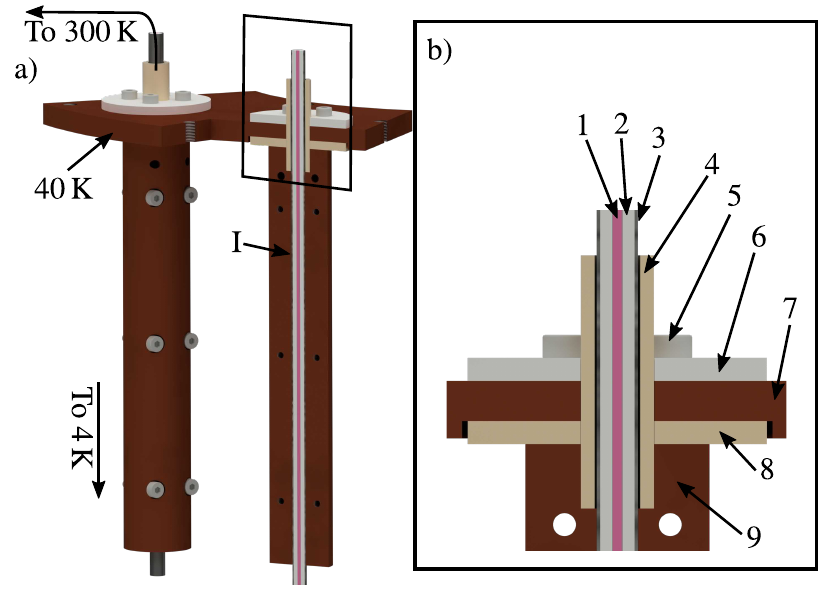}
	\caption{Thermal anchoring and electrical isolation of microwave conductors relative to the first cooling stage. a) CAD drawing of a pair of thermal anchors for high-frequency wiring. Left: outer view, right: cross section. I: Coax-line, greased with Apiezon N and pressed between two copper half-cylinders (length 10\,cm). b) magnified cross section; 1: 0.812\,mm-diameter Ag-coated BeCu center conductor, 2: 2.985\,mm outer diameter PTFE dielectric, 3: 3.581\,mm outer diameter 304 stainless steel shield, 4: Shapal tube, 5: Mounting bolts, 6: PTFE spacer, 7: 40\,K stage, 8: Shapal spacer, 9: OFE Cu half-cylinder}
	\label{fig:coax}
\end{figure}

According to the data sheet, the cable has an insertion loss of 0.83\,dB/m at 1\,GHz. With equation~\ref{thermal_conductivity_integral} and thermal conductivity data from NIST's Cryogenic Technology Group~\cite{noauthor_cryogenic_nodate}, we calculated that, in an ideal case of perfect thermal anchoring on the 40\,K stage, the heat load on the 4\,K stage will be 3.4\,mW due to thermal conduction. Figure~\ref{fig:coax} shows the thermal anchoring of the semi-rigid cable to the first cooling stage. 
We used SMA connectors (Huber+Suhner, 11\_SMA-50-3-15/111\_N, rubber gaskets removed), soldered to both ends of the semi-rigid cable using 315-LF-SOLDER-1M solder from Allectra and Castolin 157 flux. We were not able to use this type of semi-rigid cable all the way from room temperature to 4\,K. Flexible cable is required on the inside of the room temperature floating SMA feedthroughs (outer conductor not connected to flange) because of the need to close the flange after attaching the cable. The semi-rigid cable is too stiff to support the required motion. The outer conductor of a floating SMA feedthrough is not electrically connected to the flange in which it is welded in and therefore prevents ground loops. Since the first cooling stage has ample cooling power, the main concern is the heat load on the second stage. We use a 25\,cm long KAP50 cable from Allectra with SMA connectors on each end to connect the room temperature feedthroughs to the semi-rigid cable, which runs only vertically inside the radiation shield. Connecting the semi-rigid cable to the inner chamber was also not feasible due to its stiffness. We use a 3\,cm long KAP50S piece from Allectra with SMA connectors on both ends to connect the UT-141B-SS cable and the inner chamber. Floating SMA feedthroughs are used to guide the signal into the inner chamber. Inside the latter, a 10\,cm long KAP50S piece with an SMA connector on one end and an MCX connector on the other end connects to the printed circuit board (PCB) to which the ion trap is bonded. All cables used up to this point have 50\,$\Omega$ impedance. 

We measured the input return loss from the outer vacuum flange to the trap, including the PCB's coplanar waveguide, bonding wires and the trap chip to be $-2.2\pm 0.1$\,dB at 1.083\,GHz. This measurement has been carried out with a vector network analyzer (Rohde \& Schwarz ZNB8) while the system was cooled down to 5\,K. We have independently measured the insertion loss of all combined cables leading to the PCB at room temperature and at 1.083\,GHz to be $-1.16$\,dB. At lower temperatures this number should be smaller. On the other hand, the coplanar waveguide, bonding wires and trap introduce additional losses. The reflected signal will undergo these losses twice. This leads to the measured $-2.2$\,dB or 39.5\% power dissipation. This is the only variable source of heating in our system. Adding up all static heat sources, low-frequency lines, semi-rigid cables and thermal radiation, we reach a static thermal load of 165\,mW. With a cooling power of about 1.5\,W at 5\,K, this leaves us with an acceptable power dissipation of 1.335\,W which leads, in the worst case scenario of all power being dissipated on the 4\,K stage, to an acceptable average microwave input power of $\approx$3.4\,W. To verify this number, we set the temperature controller that is connected to the second stage to 5\,K, and measured the amount of heating that it puts out versus the CW microwave power we applied to the sideband electrode (labeled MWM in figure \ref{fig:trap_electrodes}) at 1.084\,GHz. We ramped up the output power of our microwave source until the temperature controller did not need to heat the system to keep it at 5\,K. We then measured the microwave power with a spectrum analyzer (Rohde \& Schwarz FPC1000/FPC-B2) and a 40\,dB attenuator (Mini-Circuits BW-N40W50+) to be $4.0\pm1.0$\,W. This is in good agreement with the calculated value. For the implementation of multi-qubit gates, power would not be applied continuously, but only during gate operation. Assuming the pulse parameters of~\onlinecite{zarantonello_robust_2019}, it would be possible to apply peak powers of $\approx$64\,W, to be compared to the 11\,W used in~\onlinecite{zarantonello_robust_2019}, before reaching cooling power limits. For future traps, it may be desirable to include proper impedance matching between the microwave feed line and the microwave conductors in the trap, which would further reduce Joule heating. 

\subsubsection{Radio frequency trap drive}
\begin{figure}
	\centering
	\includegraphics{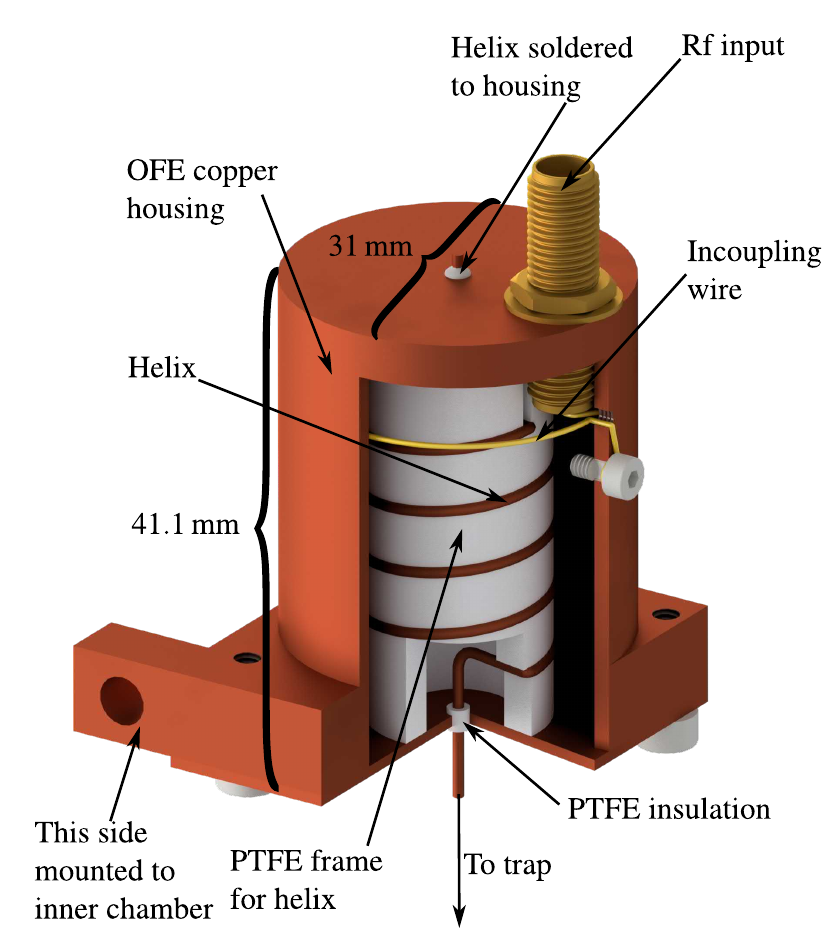}
	\caption{CAD drawing of the rf resonator (see text).}
	\label{fig:resonator}
\end{figure}

The surface-electrode Paul trap used here is operated at a radio frequency between 90 and 110\,MHz for radial confinement of the ions. We use an HP 8640B oscillator to generate the desired rf drive. A self-built helical resonator is mounted to the outside of the inner vacuum chamber and provides the required voltage step-up. The cable configuration between room temperature and the inner vacuum chamber is the same as for the microwave lines, except that the drive line is attached directly to the resonator input, instead of a feedthrough on the inner vacuum chamber. The resonator output is connected to the center pin of an SMA feedthrough on the inner chamber with a bare wire. On the inside, another bare wire continues to connect it to the PCB to which the trap is bonded. The capacitive load of a coaxial line would pull the resonator frequency too much. Figure~\ref{fig:resonator} shows an image of the rf resonator.
Following references \onlinecite{macalpine_coaxial_1959,fisk_helical-resonator_1976,siverns_application_2012}, we chose a compact cylindrical resonator of 42.1\,mm height and 31\,mm diameter with a resonance frequency of 100\,MHz. The helix is made of 1\,mm diameter OFE (oxygen free electronic) copper wire, has a pitch of 5.75\,mm, diameter of 16\,mm, height of 23\,mm and is wound on a PTFE cylinder with a 0.5\,mm deep helix-bevel for guidance. The signal is coupled in via an SMA bulkhead female-female feedthrough. We removed the isolation from the tip of a varnish insulated 0.5\,mm diameter copper wire with a rotating blade wire stripper (ISOL-EX Typ 02) and soldered a male SMA pin to it (solder: 315-LF-SOLDER-1M, Allectra). The latter is connected to the inner conductor of the SMA feedthrough and wound around the helix once, with the varnish touching the inner wall of the resonator housing. It is then guided through a 2\,mm hole to the outside of the resonator housing and attached with an M2 screw. To make small adjustments to this incoupling wire, we loosen the screw, carefully push/pull the wire from the outside and tighten it again. 
\begin{figure}
	\centering
	\includegraphics{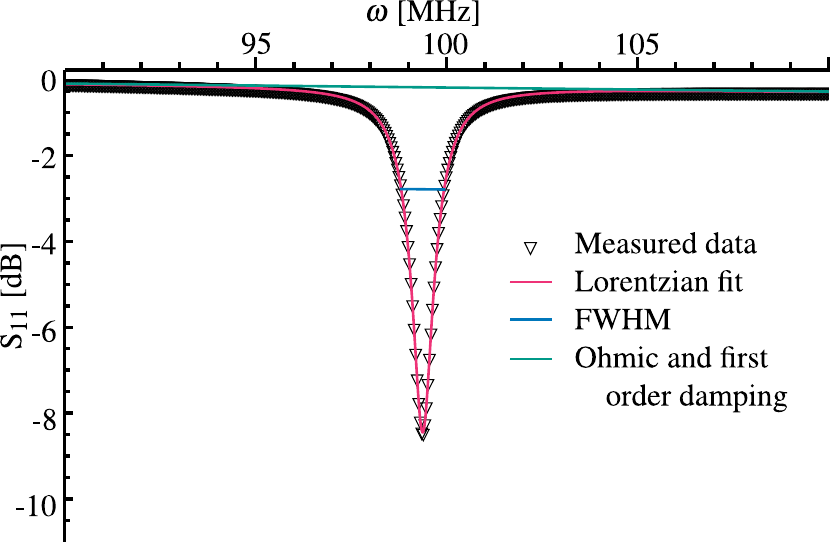}
	\caption{Measurement of the resonator's S$_{11}$ parameter, at 5\,K, while connected to the ion trap. For clarity, only every 4th data point is plotted. The center frequency is at $99.372$\,MHz, the return loss on resonance is $-8.047(4)$\,dB, the FWHM is 1.13\,MHz, the loaded Q-factor is 88.315(13). Note that because of the smaller frequency, the off-resonant return loss is different from the 2.2\,dB reported for the test of the high-frequency lines at 1.083\,GHz.}
	\label{fig:resonance}
\end{figure}

We determine the loaded Q-factor as follows: While everything is connected and cooled down to 5\,K, we measure the reflected power or S$_{11}$ parameter vs frequency by connecting a vector network analyzer to the appropriate room temperature SMA feedthrough. The resulting resonance curve is shown in figure~\ref{fig:resonance} and contains not only the frequency-dependent behavior of the resonator but also of the feed-lines. The lines have an Ohmic loss and a frequency-dependent damping, approximated to first order around the resonator's resonance frequency. We model the resonator as a series RLC circuit (impedance $Z_{res}=i \omega L + \frac{1}{i \omega C} + R $) attached to a transmission line with characteristic impedance $Z_0$ fed from a source with the same impedance $Z_0$. The voltage reflection coefficient is given by~\cite{pozar_microwave_2009} $\Gamma=\frac{Z_{res}-Z_0}{Z_{res}+Z_0}$, and the power reflection coefficient by $\left|\Gamma\right|^2=\textnormal{S}_{11}$. We find 
\begin{equation}
\textnormal{S}_{11}(\omega)=10\log_{10}\left(\frac{(\gamma - \gamma_s)^2 \omega^2 + (\omega^2 - \omega_0^2)^2}{(\gamma + \gamma_s)^2 \omega^2 + (\omega^2 - \omega_0^2)^2} \right)-\Omega \omega - R_0
\end{equation}
where an additional term has been included to correct for the frequency dependent attenuation of the line between the resonator and point of the return power measurement. $R_0$ the Ohmic loss of the feed-lines and $\Omega$ the first-order frequency dependent loss coefficient of the feed lines.
Here, $\gamma=\frac{R}{L}$ determines the frequency width of the unloaded resonator, $\gamma_S=\frac{Z_0}{L}$ is related to the source and line impedance and $\omega_0=\frac{1}{\sqrt{LC}}$ is the resonance frequency. Using a nonlinear least-squares fit to the data (see Fig.~\ref{fig:resonance}), we find $\gamma_S=2\pi \cdot 0.33984(17)$\,MHz, $\gamma=2\pi \cdot 0.78536(5)$\,MHz and $\omega_0=2\pi \cdot 99.372$\,MHz. The return loss on resonance is $-8.047(4)$\,dB. 
The loaded $Q$-factor is given by
\begin{equation}
Q=\frac{\omega_0}{\gamma+\gamma_s}.
\label{eq:q}
\end{equation}
We find $Q=88.315(13)$. In operation, we run the resonator at +23\,dBm input power. Input and reflected power are monitored continuously using a bidirectional coupler (Mini-Circuits ZFBDC20-62HP-S+) and two power detectors (Mini-Circuits ZX47-20LN-S+). 

\section{Vacuum system}
Designing vacuum systems for room temperature UHV operation requires carefully choosing materials with respect to outgassing performance and maximum baking temperature. For cryogenic vacuum systems, these requirements can be relaxed. A vacuum chamber that reaches only a moderate vacuum of about $10^{-6}$\,mbar at room temperature can, when properly cooled down, reach XHV conditions.

\subsection{Outer chamber: isolation vacuum}
\label{outer chamber}
The outer vacuum chamber has been designed to fulfill all requirements for UHV operation, but has not been baked because it only serves as an isolation vacuum for the self-contained inner chamber. Should the chamber need to be baked at some point, the use of Apiezon N grease would have to be reconsidered. We use CF flanges at every connection and currently rely on Viton\textregistered\,\,gaskets at ports that needed to be opened for trap exchanges. Annealed copper gaskets are used for seals on all other ports. Otherwise, the list of materials in the outer chamber is: OFE-copper, beryllium-copper, stainless steel, PTFE, gold-plated bronze, constantan, PEEK, Kapton, Stycast 2850FT as well as fused silica glass and optical coatings for various wavelengths on viewports. After assembling and cleaning, the chamber is evacuated with a turbomolecular pump connected to a backing scroll pump. We monitor the pressure in the outer chamber with an active Pirani/cold cathode gauge (Pfeiffer PKR 251). After about 5\,h, it reaches $10^{-6}$\,mbar and we turn on the cryocooler. When the temperature reaches 5\,K after about 10\,h, the isolation vacuum reaches less than $3\cdot 10^{-8}$\,mbar and we close the valve connecting chamber and turbomolecular pump. The vacuum is then sustained by the cryo-pumping effect. 

\subsection{Inner chamber: 4\,K ion trap housing}
\label{sec:inner_chamber}
\begin{figure}
	\centering
	\includegraphics{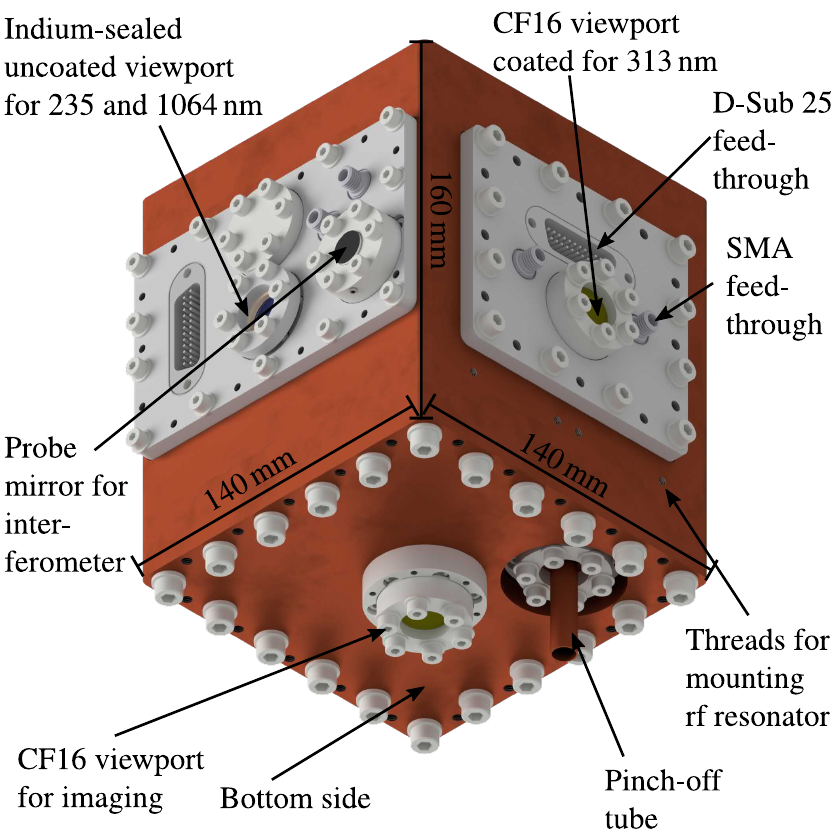}
	\caption{CAD drawing of the outside of the inner chamber. Copies of the right flange are also situated on the hidden two sides of the chamber. The left flange houses an additional port for interferometric readout of vibrations and a currently unused auxiliary CF16 port. The port for the pinch-off tube is recessed into the bottom lid because the radiation shield is so close to the bottom that the tube would touch it otherwise. The labeled laser access ports have a corresponding port on the opposite side to let the laser beam pass through the chamber.}
	\label{fig:pillbox_outside}
\end{figure}

\begin{figure}
	\centering
	\includegraphics{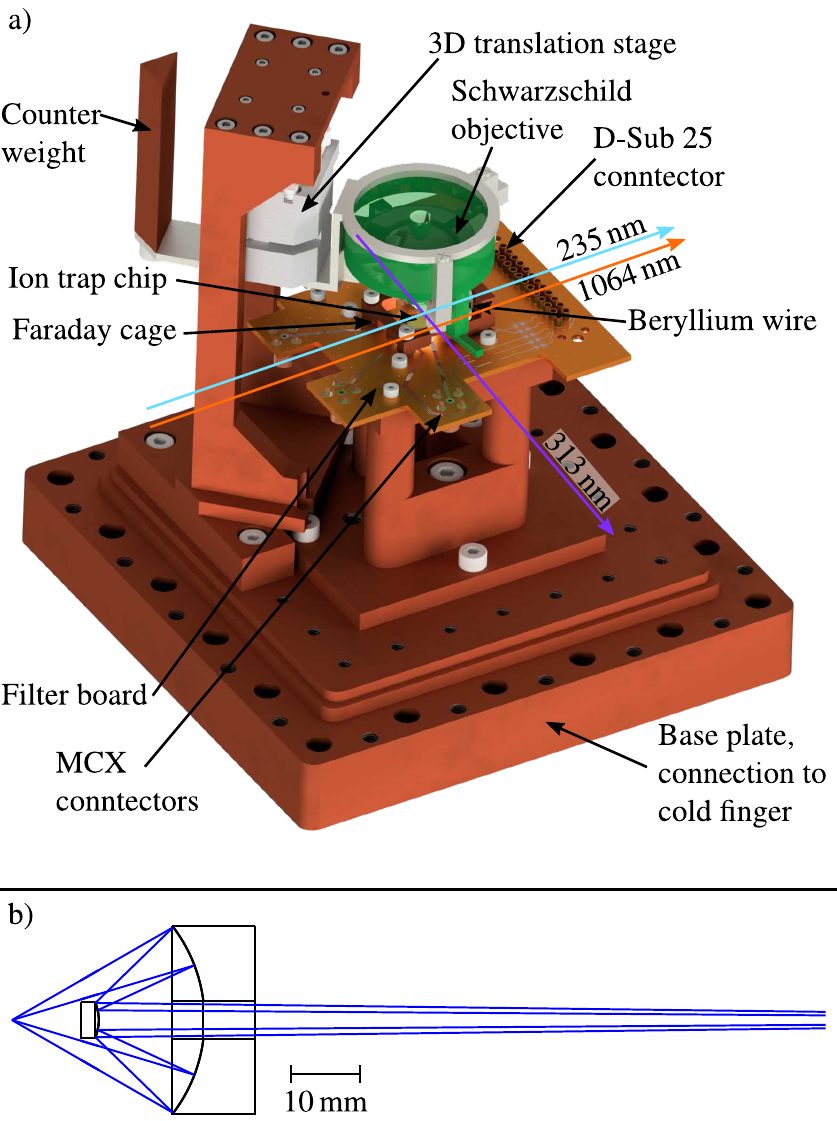}
	\caption{a) CAD drawing of the inner chamber's inside components (upside down). Electrical wiring is left out for clarity. Wires connect the lower side ot the filter board to feedthroughs in the walls (not shown either).
	b) Cryogenic Schwarzschild objective layout for ion resonance-fluorescence imaging}
	\label{fig:pillbox_inside}
\end{figure}

The inner chamber is made predominantly from OFE copper and non-magnetic stainless steel (316LN or 1.4429). It has a square $140 \times 140$\,mm$^2$ footprint and a height of 160\,mm. The OFE copper supporting structure consists of two 25\,mm thick top and bottom lids as well as a 15\,mm wall thickness enclosing body. Structurally, it would have been possible to use thinner materials, which would have the advantage of lower heat capacity and therefore faster cooldown. However, a higher heat capacity acts as a low pass filter for thermal fluctuations. Thick walls also provide better shielding from outside electric and magnetic noise.
Figure \ref{fig:pillbox_outside} shows an image of three sides of the inner chamber. 
The inner chamber's body has holes and pockets to house stainless steel flanges which contain welded-in electrical feedthroughs as well as CF16 compatible flange interfaces and spaces for 20\,mm diameter indium sealed viewports. We use a 1\,mm thick indium wire between the inner chamber copper wall and the stainless steel flange to make a vacuum tight seal. For loading, cooling and detection, we require a pulsed ablation laser beam, a photoionization beam at 235\,nm to ionize neutral $^9\mathrm{Be}$ atoms, and 313\,nm light for cooling and detection of the $^9\mathrm{Be}^+$ ions. The laser beam geometry is illustrated in Fig.~\ref{fig:pillbox_inside}. The 1064\,nm ablation and 235\,nm beams are offset from each other by 6\,mm. On the inner chamber, we use uncoated fused silica disks that are indium sealed to the adapter flange for these two beams. Both beams share one such window on either side of the chamber. For the 313\,nm cooling and detection laser beam line, we use anti-reflection coated CF16 viewports. The bottom lid accommodates two indium sealed stainless steel adapters that have CF16 interfaces. One of these is used for a CF16 viewport for ion fluorescence imaging; the other one is used for a pinch-off tube. The top lid only has M4 threads to mount it to the cold finger. The RF resonator is mounted to one of the side walls. 
An image of the chamber's inside is shown in Fig.~\ref{fig:pillbox_inside}a), with the wiring left out for clarity. The inside dimensions are $110 \times 110 \times 107$\,mm$^3$. At the center, there is a copper block of $49 \times 49 \times 7.9$\,mm$^3$, connected with a series of mechanical adapters to the base plate attached to the cold finger. At the center of this copper block is a 1.63\,mm tall plateau with a surface area of  $\approx 4.8 \times 4.8$\,mm$^2$. An ion trap chip of the same area (sapphire substrate, thickness $635\,\mu\mathrm{m}$) is glued with an EPO-TEK H74 (EPOXY TECHNOLOGY, INC.) thermally conductive epoxy to the plateau. A PCB (`filter board'), made out of gold coated nominally $1.52\pm 0.11$\,mm thick Rogers 4350B material (CONTAG AG) with a $5 \times 5$\,mm hole surrounds the trap and is bolted onto the mounting block. The PCB accommodates one 25 pin D-sub connector for low-frequency signals as well as four MCX connectors for RF and microwave signals. The low-frequency signals pass an RC lowpass (1\,k$\Omega$ Anaren R1A15081001J3A0 and 820\,pF Novacap 0603N821J101P). The capacitors are chosen for their NPO dielectric~\cite{teyssandier_commercially_2010}. The microwave signals are guided towards the trap with a 50\,$\Omega$ coplanar waveguide. We left a 1.05\,mm gap between the ground plane and the 0.4\,mm wide rf line to minimize the capacitive load on the resonator. Wire bonds establish connections between the filter board and the ion trap for all electrical lines. A Faraday cage with ports for laser beams and at least 1\,mm away from electrical lines surrounds the ion trap. It has an opening for imaging light to pass through, which is covered with a gold mesh (Precision Eforming MG17) to keep a defined ground potential above the trap. The Faraday cage structure also houses a 0.5\,mm diameter beryllium wire, located 13\,mm away from the trap center. An opening of 2.4\,mm diameter gives way for the ablation laser to hit the Be wire. Another opening of 1.4\,mm diameter placed between the wire and the trap center acts as a guide for ablated Be atoms to reach the trap. The 1.4\,mm hole also acts as an aperture, minimizing Be contamination of the chip.

A Schwarzschild objective is placed 8\,mm above the trap for ion detection through resonance fluorescence imaging. It is mounted on a 3D translation stage made out of a stack of two x-nanopositioners (SLC-1720-CR-UHVT-NM-TI, SmarAct GmbH) and one z-nanopositioner (ANPz101, attocube systems AG), all fully non-magnetic. The six current-carrying lines needed to connect the translation stage are connected to one of the 25 pin D-sub ports of the inner chamber (not the one used for trap voltages). On the z-positioner, we mounted a titanium structure that holds the objective and an OFE-copper counter weight to minimize torque on the positioner. With 96\,g, this combination is only 4\,g lighter than the maximum specified weight of the attocube ANPz101. To improve cryogenic vacuum, 1\,mm holes were drilled in three $1 \times 2 \times 1$\,cm$^3$ OFE copper boxes filled with pellets of activated charcoal which, due to its high surface to volume ratio, provides an additional absorption effect. 

We assemble the inner chamber in a cleanroom. After each step, we test all electrical connections. We close the chamber and transport it to the lab. In case it becomes necessary to open the inner vacuum chamber in the laboratory, a small cleanroom tent is available to be assembled next to the optical table. We connect the pumping station (see section \ref{outer chamber}) to the pinch-off tube, connected to one of the bottom CF16 ports and pump until the vacuum reaches below $6\times 10^{-6}$\,mbar, measured directly before the pinchoff tube leading to the chamber. The vacuum does not reach lower pressures, mostly because of the charcoal's outgassing and desorption at room temperature. The pressure inside the vacuum chamber is expected to be higher due to the small conductance of the pinch-off tube, but cannot be measured. We then pinch off the inner chamber and mount it to the outer chamber. We do not bake the inner chamber prior to cooling down. In order to allow for substantial baking, the indium seals would have to be replaced with gold seals, with a final baking temperature limited by the PEEK connectors to about 250\,$^\circ$C. We close the outer chamber and proceed as described in section \ref{outer chamber}. After over-night pumping, the inner chamber will be at 5\,K. Since the inner chamber offers no space for a vacuum gauge, we rely on ion-based measurements to characterize the background pressure. If the ions are being Doppler cooled, there is a probability that they will react with background molecular hydrogen, as long as they are in the excited state: Be$^+ (^2 \textnormal{P})$+H$_2 \rightarrow$ BeH$^+$+H. At 5\,K and with Doppler cooling on, according to previous work, this is the dominant loss channel for $^9$Be$^+$ ions in the trap\cite{sawyer_reversing_2015}. The reaction rate $k_L$ can be described by a Langevin model and $k_L=1.6\times 10^{-9}$\,cm$^3$\,s$^{-1}$ has been determined previously\cite{roth_ion-neutral_2006}.
The H$_2$ pressure can be deduced from the ion reaction rate $\tau^{-1}$ of Be$^+$ at a given cooling laser intensity, the Langevin model and the ideal gas law\cite{sawyer_reversing_2015}:
\begin{equation}
P_{\textnormal{H}_2}=k_B  \rho_{\textnormal{H}_2} T\,,
\label{eq:pressure_calc_1}
\end{equation}
where $P_{\textnormal{H}_2}$ is the H$_2$ pressure, $k_B$ the Boltzmann constant, $\rho_{\textnormal{H}_2}$ the H$_{2}$ number density and $T$ the temperature. $\rho_{\textnormal{H}_2}=(k_L \tau')^{-1}$ is related to the effective ion lifetime $\tau'$ and to the Langevin reaction rate $k_L$, with $\tau'=\tau/p(^2 \textnormal{P})$, where $\tau$ is the observed lifetime and $p(^2 \textnormal{P})$ the probability of the ions being in the excited state. We have not observed ion loss due to background gas collisions so far. We therefore cannot deduce a background pressure from our loss rate. In the early trials of trapping and commissioning the experiment, we used about 5 times the saturation intensity to cool the ions, which leads to a near 50\% occupation of the $^2$P state,  $p(^2 \textnormal{P})=0.5$. As in~\cite{sellner_improved_2017}, we model the probability that an H$_2$ molecule actually collides with an ion as a Poisson process: $f(n;\lambda)=\lambda^n \textnormal{exp}(-\lambda)/n!$ with $\lambda=t_{exp}/\tau'_{lower}$ and $n_0=0$ events, where $t_{exp}$ is the accumulated time that trapped ions were exposed to the background gas and $\tau'_{lower}$ is the lower limit on the ion lifetime. At a chosen confidence level CL, the following equation holds\cite{sellner_improved_2017}:
\begin{equation}
\textnormal{CL}=1-\epsilon=\sum_{n=n_0+1}^{\infty} f\left( n;\frac{t_{exp}}{\tau'_{lower}} \right)
\label{eq:pressure_calc_2}
\end{equation}
\begin{equation}
\rightarrow \epsilon=\sum_{n=0}^{n_0} f\left( n;\frac{t_{exp}}{\tau'_{lower}} \right).
\label{eq:pressure_calc_3}
\end{equation}
We have accumulated about 15 days of equivalent one-ion lifetime: $t_{exp}=15$\,d. Solving equation~\ref{eq:pressure_calc_3} for $\tau'_{lower}$ and inserting the result into equation \ref{eq:pressure_calc_1} yields a maximum background pressure $P_{\textnormal{H}_2}=1.2\times 10^{-12}$\,mbar at a confidence level of 68\% and $P_{\textnormal{H}_2}=2.5\times 10^{-12}$\,mbar at CL=90\%.

\section{Vibration isolation}

\subsection{Design}
\begin{figure}
	\centering
	\includegraphics{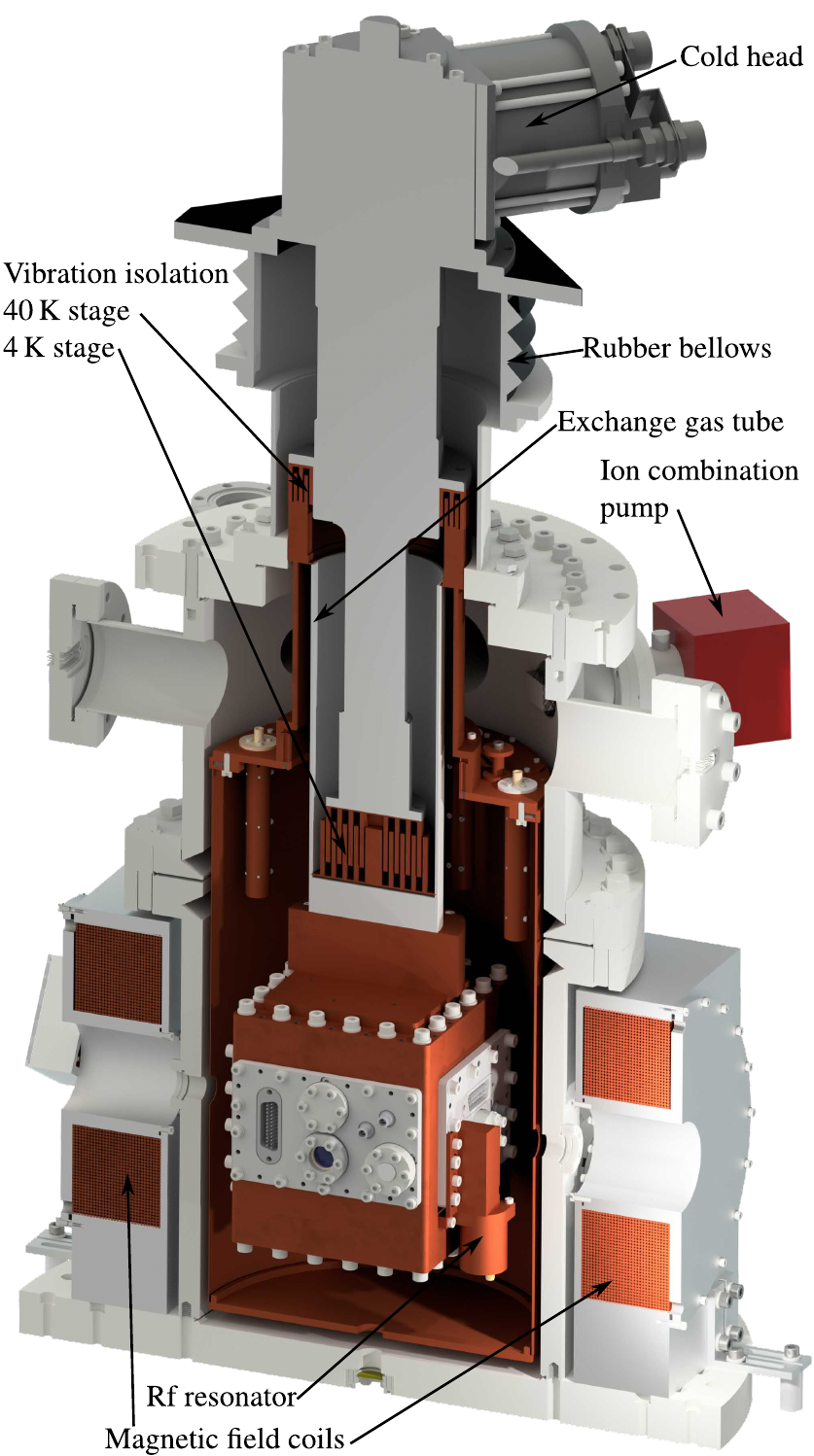}
	\caption{CAD drawing of the outer chamber, radiation shield and ultra-low-vibration interface with non-cut inner chamber attached. Wiring was left out for clarity. The cold head is mounted to the lab ceiling with the device shown in figure \ref{fig:cryo_mount}, which is omitted in this picture. Everything else is mounted to the optical table (also see figure~\ref{fig:vacuum_system_cut}).}
	\label{fig:vacuum_system_cut}
\end{figure}
\begin{figure}
	\centering
	\includegraphics{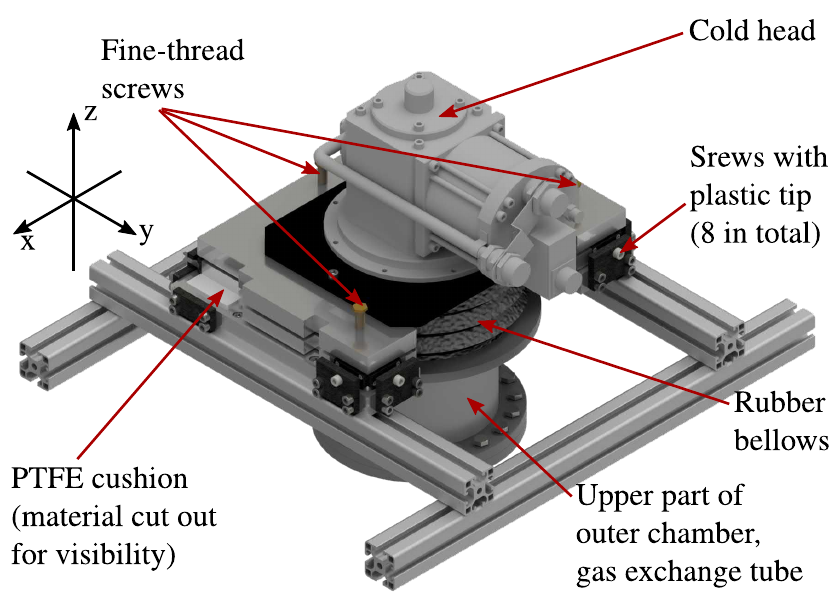}
	\caption{CAD drawing of the alignment platform. The platform is mounted to the lab ceiling on aluminum profiles (not all shown for clarity). Three fine-thread screws determine the relative angle between optical table and cold head as well as z-position. Eight screws with plastic tips adjust the x-y movement. The plastic, together with the PTFE cushions, on which the two upper plates glide, assures electrical isolation of the cold head and the support structure, reducing ground loops. The rubber bellows are the only mechanical connection between the cold head and the outer vacuum chamber.}
	\label{fig:cryo_mount}
\end{figure}

The cold head of a GM cooler contains moving parts which induce significant vibrations, on the order of 100\,$\mu$m at the cold finger\cite{ekin_experimental_2015} to which the experiment is attached. In the present system, shown in figure \ref{fig:vacuum_system_cut}, the Sumitomo RDK-415D cold head is decoupled from the rest of the experimental setup with a buffer gas vibration isolation system manufactured by ColdEdge Technologies. The vibrating part is not mounted to the optical table, on which the ion trap and lasers are mounted, but attached to the lab ceiling. The cold finger is fitted into a tube with a helium exchange gas which transmits the cooling power without the vibrations. Rubber bellows on top of this tube constrain the helium and are the only mechanical connections between the vibrating cold head and the gas-filled tube. At the bottom of the tube, helium is at the cold finger temperature, while at the top it is at room temperature. Some helium leaks through the bellows; it is therefore constantly replenished. We reduce the pressure of a 50\,L and 200\,bar cylinder of 99.999\% pure helium using two stages of pressure regulators to about 30\,mbar above ambient pressure and feed the gas into the top end of the exchange gas tube. The overpressure keeps non-helium gases out, especially during cool down as the gas is contracting. While the cryocooler is running, the consumption is about 3 cylinders per year. The exchange-gas tube is part of the outer vacuum chamber, which is mounted on the optical table. In the middle of the tube, where the first cooling stage of the GM cooler is located, we mount the radiation shield. At the tube's bottom, we connect the inner vacuum chamber. On both temperature stages, the vibration isolation works as follows: concentric copper cylinders a few cm long and about 5\,mm thick are connected to the cold finger. In the gaps between the cylinders, similar concentric copper cylinders are mounted to the exchange-gas tube. There is a gap of about 0.5\,mm between these matching pairs of cylinders. One has to ensure that the rings do not touch, as this would enable the direct transmission of mechanical vibrations. The cold head therefore has to be aligned relative to the exchange-gas tube with better than 0.5\,mm accuracy. We have built an alignment platform, mounted to the lab ceiling, on which the cold head rests, with three contact points on an x-y-translation stage to align the cold head relative to the optical table. Figure \ref{fig:cryo_mount} shows the alignment platform. 
To minimize vibrations coupling in from the floor, the optical table is supported by pneumatic legs outfitted with repositioning valves with a positioning accuracy of better than 0.3\,mm (86-19888-02, TMC) so that, after movement, the optical table comes back to a position in which the vibration isolation does not need to be realigned. In principle, it should be possible to check for electrical contact between the cold head and the outer vacuum chamber to ensure that the copper cylinders do not touch. In reality, metallic helium lines and electrical cables are connected to the cold head, which, via the helium compressor, are grounded, like the optical table. Checking for electrical contact is therefore only reliable if the helium lines are disconnected. Reconnecting them requires torque and can misalign the cold head again. Instead, we measure the inductance, essentially the length of a current loop, between cold head and optical table. This can be done while everything is connected and even while the cooler is running. We see a change in inductance by two orders of magnitude from $10^{-7}$\,H to $10^{-5}$\,H depending on whether the copper cylinders are touching or not. Alignment takes about 5\,min and has to be redone about once in two weeks. In case of pressurized air supply failure, the optical table settles down by about 3\,cm. The rubber bellows between the cold head and the outer chamber are then stretched by that amount. We made sure that they are long enough to endure this. Also, the concentric copper cylinders are pulled away from each other vertically. It may be that for non-ideal vertical alignment, which is hard to detect because it will only marginally reduce cooling power, the 3\,cm movement is enough so that the cylinders are completely separated in the vertical direction. When the pressure comes back, the table moves upwards with a superimposed sheering movement; as a result, the cylinders might chock and the cold head could be lifted out of its mount. To prevent this, an interlock monitors the air pressure and prevents the return of pressurized air after pressure loss so that the system can be brought back in position while under observation. 

\subsection{Characterization}
We use a Michelson-type interferometer to characterize the residual vibrations of the inner vacuum chamber relative to the optical table. The interferometer uses 626\,nm light from the sum frequency generation setup (see section \ref{sec:laser}). The 1050\,nm and 1550\,nm lasers have a specified short-term linewidth of less than 10\,kHz and less than 1\,kHz, respectively. The probe mirror is mounted to the inner chamber, while the reference mirror is on the optical table, supported by a piezo actuator. The arm lengths were equal to within 5\,cm. For the measurement of movement parallel to the optical table, we placed the interferometer on a breadboard surrounding the outer chamber (h in figure \ref{fig:whole_system_cut}). For a second measurement, along the direction of gravity, we placed it on a smaller breadboard below the optical table, usually reserved for ion imaging optics and camera (f in figure \ref{fig:whole_system_cut}).

\begin{figure*}
	\includegraphics{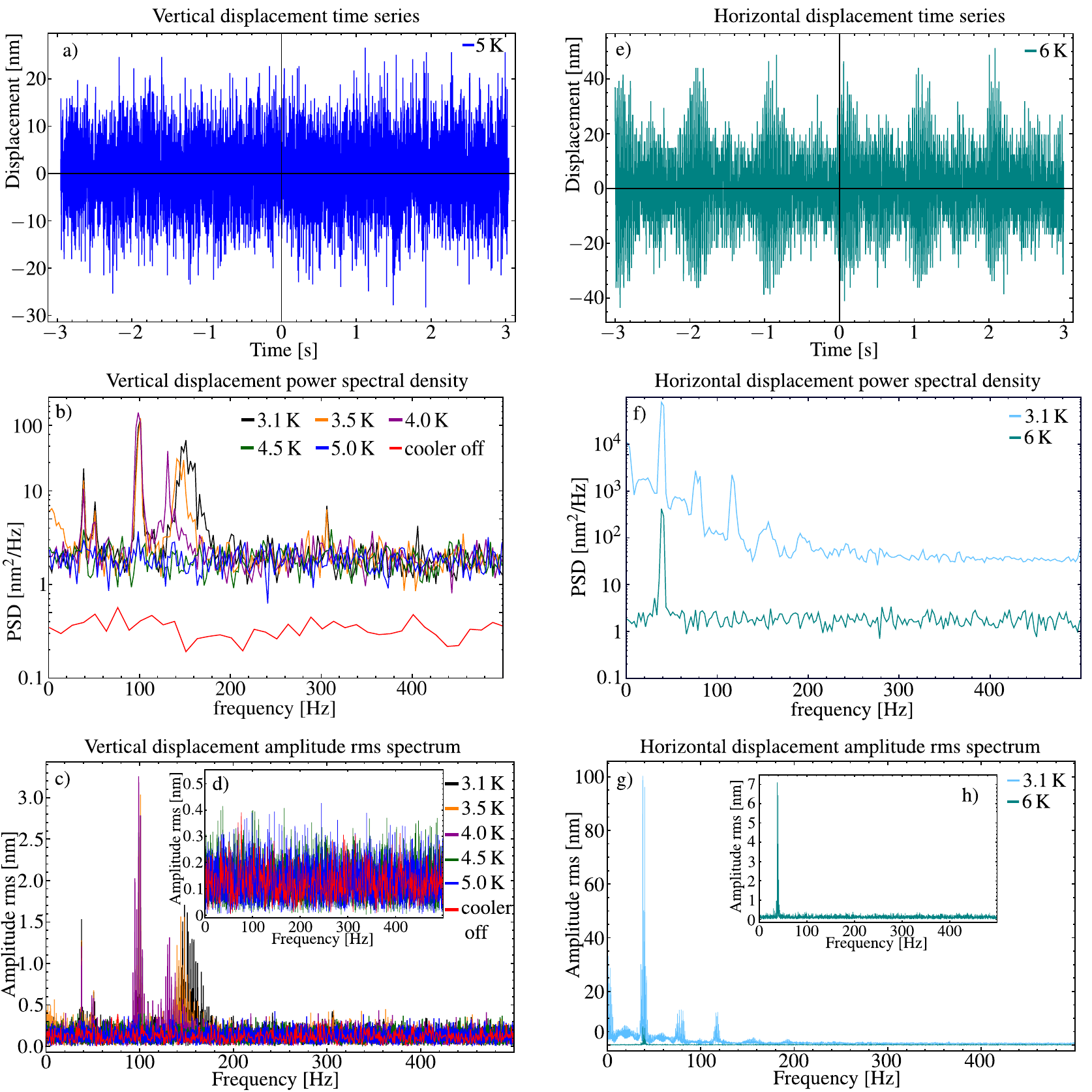}
	\caption{Vertical and horizontal vibration time series, resulting power spectral density (PSD) and vibration amplitude spectra.
		a) Time series for vertical vibrations at 5\,K.
		b) Comparison of PSD spectra for different second stage temperatures and the case of a shut-off cold head. Cooling was shut off at 5\,K and the measurement was taken about 1\,min afterwards. The peak at around 40\,Hz is the eigenfrequency of the upper breadboard (see h in figure \ref{fig:whole_system_cut}) surrounding the vacuum chamber. The peak at around 100\,Hz is the eigenfrequency of the lower breadboard (see f in figure \ref{fig:whole_system_cut}). For this measurement, the interferometer was placed on the lower breadboard. The data was taken with a RBW of 0.167\,Hz. For clarity, we integrated over 15 data points and the RBW of the graph is 2.5\,Hz.
		c) Comparison of vibration amplitude rms spectra for different second stage temperatures and the case of shut-off cold head. Same data source as b), converted to rms with a resolution bandwidth of 0.167\,Hz.
		d) Same as c) but showing only data for 4.5\,K and above, where no liquid helium is present.
		e) Time series of horizontal vibrations at 6\,K. We attribute the presence of the signals, recurring at about 1\,s intervals, not to a failure or misalignment of the ultra-low-vibration interface but to acoustic excitation of the upper breadboard by the sounds emitted by of the cold head operating at a rate of about 1\,Hz. A more rigid breadboard can be desirable in the future. Early measurements have shown one order of magnitude larger displacements which were reduced to the present level by adding more mechanical connections or posts between the upper breadboard and the optical table.
		f) Horizontal PSD spectra for 3.1\,K and 6\,K. The prominent peak is at the eigenfrequency of the upper breadboard surrounding the vacuum chamber (see h in figure \ref{fig:whole_system_cut}), on which the interferometer is placed. Harmonics are present at around 80\,Hz and 120\,Hz. The data was taken with a RBW of 0.167\,Hz. For clarity, we integrated over 15 data points and the RBW of the graph is 2.5\,Hz.
		g) Horizontal vibration amplitude rms spectra for 3.1\,K and 6\,K. Same data source as f), converted to rms with a resolution bandwidth of 0.167\,Hz.
		h) Same as g) but only at 6\,K.
	}
	\label{fig:vibrations}
\end{figure*}

We detected the signal on one port of a differential photodetector (2007 Nirvana, Newport). The second port was illuminated with laser light that was split off from the laser source before entering the interferometer. We adjusted the respective power levels to obtain a signal centered around 0\,V. We locked the signal to the zero-crossing with a PID controller (PID 110, Toptica) with feedback on the piezo actuator, therefore forcing the reference mirror to exhibit the same movement as the probe mirror. We recorded the piezo voltage on an oscilloscope (HMO 3524, Hameg) and calculated the corresponding travel distance. The voltage to distance relation of the piezo was calibrated by replacing the probe mirror with a mirror on the optical table and scanning over four fringes. We measured vibrations at different temperatures around the boiling point of helium. At temperatures where the helium in the exchange tube is liquefied, zero to peak vibration amplitudes of about 50\,nm in the vertical direction and about 1.5\,$\mu$m in the horizontal direction were observed. When the helium was gaseous, we measured vibration amplitudes of 29\,nm (7.8\,nm rms) in the vertical direction and 51\,nm (13.5\,nm rms) in the horizontal direction. Figure~\ref{fig:vibrations} shows time series, power spectral density (PSD) and vibration amplitude rms spectra. The proper way to report such data as a function of frequency is the power spectral density; we also report the vibration amplitude rms spectra for comparison to parts of the literature. 

For comparison, consider the system described in reference \citenum{pagano_cryogenic_2019} based on a cryogenic ion trap mounted to a GM cooler, mechanically decoupled via an exchange gas, comparable to our system. The vibration amplitudes in \ref{fig:vibrations}a) are roughly a factor of two smaller than shown in figure 7c) of reference \citenum{pagano_cryogenic_2019}.

Comparing the horizontal rms spectra of figure \ref{fig:vibrations}j), obtained with a resolution bandwidth (RBW) of 0.167\,Hz, to the 7.7\,K measurement of figure 7a) of reference \citenum{pagano_cryogenic_2019}, obtained with an RBW of 0.1\,Hz, our dominant rms values are about a factor of 7 lower without correcting for the different RBWs. RBW correction would increase that factor even further. Now consider the custom vibration isolation for a pulse tube cooler described in reference~\citenum{micke_closed-cycle_2019}. Comparing the vertical rms spectra of figure \ref{fig:vibrations}d) to figure 13f) of reference \citenum{micke_closed-cycle_2019}, our system shows more than one order of magnitude lower vibration amplitudes. Comparing the horizontal rms spectra of figure \ref{fig:vibrations}h) to figure 13d) of reference~\citenum{micke_closed-cycle_2019}, our results show the same order of magnitude, with reference~\citenum{micke_closed-cycle_2019} performing slightly better at some frequencies. Again, the comparison with reference~\citenum{micke_closed-cycle_2019} (RBW 0.1\,Hz) is not corrected for RBW. Correcting for this would further improve our numbers. Reference~\citenum{brandl_cryogenic_2016-1} describes an alternative approach based on a flow cryostat. Comparing horizontal time series and rms spectra of figure \ref{fig:vibrations}e) and h) with corresponding graphs in figure 5 of reference \citenum{brandl_cryogenic_2016-1}, our system has only twice the time-series amplitude and comparable amplitudes in the rms spectrum relative to this system that does not have any moving parts at all close to the trap. Reference \citenum{brandl_cryogenic_2016-1} does not provide the RBW used in the meaurements and does not show data for vertical vibrations.

\section{Laser systems}
\label{sec:laser}
For Doppler cooling, we use light at 313\,nm, generated by sum frequency generation of two infrared lasers to the visible with subsequent frequency doubling to the UV~\cite{wilson_750-mw_2011}. We combine 5\,W of a 1550\,nm and 2.5\,W of a 1050\,nm laser, both fiber lasers (ADJUSTIK, NKT Photonics) with subsequent fiber amplifiers (BOOSTIK, NKT Photonics), inside a periodically-poled lithium niobate (PPLN) crystal to reach about 1\,W of 626\,nm light. We stabilize the frequency to an iodine line in a Doppler-free saturated absorption spectroscopy setup. We use a non-linear barium borate (BBO) crystal inside a bow-tie cavity to double the laser frequency and reach about 100\,mW of 313\,nm light. Using two 50/50 beam splitters, the light is delivered to three AOM (acousto-optic modulator) setups for switching and frequency shifting. One AOM shifts the light by -100\,MHz for far-detuned Doppler cooling. Another beam is shifted by +624\,MHz in a double-pass AOM setup for near-detuned Doppler cooling and detection. A third beam also passes a double-pass AOM and is shifted by -634\,MHz. All beams are combined again and coupled into a photonic crystal fiber~\cite{gebert_damage-free_2014} before passing through polarization optics and being focused into the vacuum chamber. The -634\,MHz detuned beam is used for repumping during cooling and state preparation sequences. Since this repumper is used only during an experimental sequence and the far-detuned Doppler beam is only used during loading, these beams are never on at the same time. We decided to circumvent the power loss on a 50/50 beam splitter by feeding the repumper through the 100\,MHz AOM and overlapping its 0th order of diffraction with the 1st order of the far-detuned Doppler beam. The other 2.5\,W of the 1050\,nm laser may be used to produce light for stimulated-Raman transitions and ground state cooling by summing with another laser source near 1550\,nm~\cite{hahn_two-qubit_2019}.

For production of neutral beryllium atoms, we send a 5\,ns pulse of a 1064\,nm laser onto a beryllium wire. The laser is a Q-switched Nd:YAG laser with a pulse energy of 28\,mJ and variable repetition rate of 1-15\,Hz (Minilite I, Continuum).

We ionize neutral beryllium via a two-photon process~\cite{lo_all-solid-state_2014}. A diode-seeded tapered amplifier (TA pro, Toptica) that runs at 940\,nm is frequency-doubled in a periodically-poled potassium titanyl phosphate (PPKTP) crystal inside a bow-tie cavity to 470\,nm. In a second frequency doubling cavity with a BBO crystal, 235\,nm light is generated. We start with 1\,W in the infrared and end up with 12\,mW in the UV, of which we use about 1.5\,mW for ionization.

\section{Imaging system}
We collect the fluorescence of the beryllium ions with an in-vacuum Schwarzschild-type objective. To be able to discriminate between different internal states of the ion, a high photon collection efficiency and hence a high numerical aperture (NA) on the input side of the optical system is required. Placing a high-NA lens for photon collection outside the cryogenic environment would have required large windows in the cryogenic shields which could introduce an additional heat load. The in-vacuum objective is shown schematically in Fig.~\ref{fig:pillbox_inside}b). It consists of a primary concave mirror with an ROC (radius of curvature) of $\approx22\,\mathrm{mm}$ and of a secondary convex mirror with an ROC of $\approx7\,\mathrm{mm}$. The secondary mirror sends the resonance fluorescence light through an opening in the primary mirror and through a window in the inner and outer vacuum chamber to produce an intermediate image just below the outer vacuum chamber, about 1.5\,cm below the optical table's top surface. Details on the construction and optimization of the objective will be reported elsewhere. In the intermediate image plane, a motorized four-blade aperture can be used to cut off stray light. The intermediate image is imaged onto the detectors by a combination of two lenses. Moving the lenses along an optical rail, magnification factors between 10 and 95 can be chosen. Using a motorized flip mirror, the light can be sent either to a photomultiplier tube (Hamamatsu H10682-210) or to an EMCCD camera (Andor iXon Ultra DU-888). If, in a future upgrade of the apparatus, the lenses are replaced by a reflective relay telescope, fully achromatic imaging can be achieved for dual species operation, such as for quantum computing. 

\section{Trap loading}
\label{sec:trap_loading}
The goal of the ablation laser beam alignment is to hit the beryllium target in such a way as to maximize the number of neutral atoms in the ablation plume making it to the trap center for subsequent photoionization. The ablation laser beam enters though the same windows as the ionization beam, parallel to the latter at an offset of 6\,mm. The target is a 0.5\,mm diameter beryllium wire. We position a camera behind the exit windows, operate the laser in low power mode and coarsely align the beam such that the shadow cast by the wire can be observed with the camera. Some of the light reflected by the wire is directed towards the trap surface and collected by the imaging system. Maximizing the amount of scattered light appears to also provide a favorable ablation beam fine-alignment for ion loading. The shape of an ablation plume is a semi-ellipsoid, with the major axis (the direction of emission) perpendicular to the surface~\cite{schou_plume_2007}. Diffuse light scattering follows Lambert's cosine law, with the strongest scattering occurring perpendicular to the scattering surface. The directions of the ablation plume and of the strongest light scattering should therefore coincide, making ablation laser light collection in the ion imaging system a valid tool to optimize the ablation beam position.

Aligning the 313\,nm and 235\,nm beams is done using a standard procedure for surface-electrode ion traps. The laser beams are aligned parallel to the chip surface and positioned such that they leave a trace of light on the surface, approximately the width of the beam waist ($\approx20\,\mu\mathrm{m}$). This trace is imaged onto the camera, allowing for the identification of known geometric trap features (gaps between electrodes). Following such features by displacing both the laser beam and the objective, the region surrounding the trap center can be imaged. For the 313\,nm beam, we identify the coordinates of features $\alpha$ and $\beta$ in figure~\ref{fig:trap_electrodes} on the camera image. The trap center is located in the middle between these two points, and the laser beam is aligned such that its trace on the trap surface as observed with the imaging system goes through both points. For the 235\,nm beam, an identical procedure is followed using reference points $\gamma$ and $\delta$. Both beams are then displaced from the surface by the expected ion-to-surface distance of $\approx70\,\mu\mathrm{m}$; the objective is also displaced vertically by the same distance to image the spot below the surface where the ions are expected to appear. The alignment procedure may lead to the emission of electrons from trap surfaces. This can be of concern because of the potential for charge buildup on exposed dielectric surfaces. However, the closest dielectrics are at the bottom of narrow gaps between thick metal electrodes~\cite{bautista-salvador_multilayer_2019}, effectively shielding the ion(s) from such charges. The effects of any such charges are therefore typically small and also mostly stationary - trap frequencies match expectations at the few percent level, and micromotion compensation settings do not change by much following laser beam realignment. 

For loading, we apply appropriate voltages on all trap electrodes that are required to establish a confining potential. The control signals $\mathrm{DC}_{1-10}$ are generated by a fast arbitrary waveform generator~\cite{bowler_arbitrary_2013}. Before entering the outer chamber, each of these signals passes through a 0.3\,MHz low pass filter (LPF-B0R3, Mini-Circuits). 
Magnetic field coils (see \ref{fig:vacuum_system_cut}) provide a quantization field of 22.3\,mT at the trap center. Each coil consists of 621 windings of 2\,mm diameter copper wire. About 36\,A of current are required, dissipating about 8\,kW of power. The heat is removed by pushing $\approx20^\circ\mathrm{C}$ cooling water through each coil housing at 10\,L/min. The reported vibration measurements have been carried out while the cooling water was running. The angle between the quantization field direction and the axial trap direction is 70$^\circ$ (see B$_0$ in figure \ref{fig:trap_electrodes}). We switch on <the 235\,nm and both Doppler beams of the 313\,nm laser. Then we send one 9.3\,mJ pulse of the 1064\,nm laser onto the beryllium target. Neutral beryllium is ablated and forms a gas jet directed towards the trap center. Atoms with low kinetic energy are ionized by 1.5\,mW of the 235\,nm laser, cooled by 300\,$\mu$W of the far detuned and 10\,$\mu$W of the near detuned 313\,nm laser and trapped in the potential as shown in the inset of figure~\ref{fig:trap_electrodes}. 

\section{Conclusion}
We have designed and commissioned an ultra-low-vibration closed-cycle cryogenic ion trap setup for surface-electrode ion traps. Vibration amplitudes at the noise level of a Michelson interferometer were found. Residual vibrations in the horizontal direction can be further suppressed using a more rigid optics platform. With of order 100 lines for low-frequency control voltages, eight rf and microwave lines and optical access along 2 orthogonal directions, the system will allow the implementation of a small-scale quantum processor or simulator with trapped ions based on integrated microwave control. 

\section*{Data Availability Statement}
The data that support the findings of this study are available from the corresponding author upon reasonable request.

\begin{acknowledgments}
We thank the IQ machine shop for much needed advice and support in building the apparatus. We thank Wissenschaftlicher Gerätebau at PTB for further support. We thank Bernhard Roth, Brian Sawyer and Terry Rufer for helpful discussions. We acknowledge funding by DFG through SFB 1227 `DQ-mat' (project A01), the cluster of excellence `Quantum Frontiers' and the European Union through the QT flagship project `MicroQC'.
\end{acknowledgments}

\bibliography{cryo_paper}

\end{document}